\documentclass[preprint,12pt]{elsarticle}

\usepackage{amssymb}
\usepackage{slashed,graphicx}
\usepackage{amsmath,amsfonts,mathrsfs}

\newcommand\g{\gamma}
\newcommand\del{\delta}

\newcommand\e{\eta}
\newcommand\q{\theta}

\renewcommand\l{\lambda}
\newcommand\m{\mu}
\newcommand\n{\nu}
\newcommand\x{\xi}

\renewcommand\r{\rho}

\renewcommand\j{\psi}

\newcommand\J{\Psi}

\renewcommand\d{\partial}

\newcommand{\lan}{\langle}
\newcommand{\ran}{\rangle}

\newcommand{\no}{\nonumber}

\newcommand{\diracslash}[1]{#1\llap{/\kern2pt}}

\newcommand{\be}{\begin{equation}}
\newcommand{\ee}{\end{equation}}
\newcommand{\bea}{\begin{align}}
\newcommand{\eea}{\end{align}}
\newcommand{\ba}[1]{\begin{array}{#1}}
\newcommand{\ea}{\end{array}}

\usepackage{amssymb}

\journal{Physics Letters B}

\begin{document}

\begin{frontmatter}



\title{Photon mass via current confinement}

\author[aa]{Vivek M. Vyas}
\ead{vivekv@rri.res.in}
\address[aa]{Raman Research Institute, Sadashivnagar, Bengaluru - 560 080, India}

\author[bb]{Prasanta K. Panigrahi}
\ead{pprasanta@iiserkol.ac.in}
\address[bb]{Department of Physical Sciences, Indian Institute of Science Education and Research Kolkata, Mohanpur - 741 246, India}


\begin{abstract}
A parity invariant theory, consisting of two massive Dirac fields, defined in three dimensional space-time, with the confinement of a certain current is studied. It is found that the electromagnetic field, when coupled minimally to these Dirac fields, becomes massive owing to the current confinement. It is seen that the origin of photon mass is not due to any kind of spontaneous symmetry breaking, but only due to current confinement.
\end{abstract}


\end{frontmatter}



\section{Introduction}
Field theories in three dimensional space time have been a subject
of intense study since a couple of decades now. There are several
reasons which make such field theories interesting. Firstly often
the theories in lower dimensions are simpler than their higher dimensional counterparts. 
Secondly, it offers new structures like
possibility of gauge invariant mass term for gauge field in the form
of Chern-Simons term in the action. Interestingly, it was recently found that planar QED with a tree level Chern-Simons term 
admits a photon which is composite \cite{abhi}. 
Theories with Chern-Simons
term are found to play an important role in physics of
quantum Hall effect and anyonic superconductors \cite{Nagaosa, panigrahi, any, wilczekbook, laughlin}. Models which
exhibit dynamical mass generation and spontaneous chiral symmetry breaking have been
constructed and extensively studied \cite{appl1,appl2,massgen1,massgen2,massgen3}. 
In recent years, with the discovery of graphene \cite{novoselov} and topological insulators \cite{zhang}
there has been a renewed interest in the study of lower dimensional field theories.

Colour confinement is one of the still not well understood aspect of
QCD. One of the main
hindrance is the fact that the low energy dynamics in such
theory becomes non-perturbative, which makes dealing with them
difficult. To circumvent this difficulty, there have been attempts
to assume colour confinement from the beginning and work subsequently to
see if one can get some idea about the dynamics of non-Abelian gauge
fields \cite{eguchi, kikkawa, sri}. In a remarkable paper by Srinivasan and Rajasekaran, it was shown that by
assuming quark confinement it was possible to get QCD out of it \cite{sri}. Confinement has also been studied in theories defined in three dimensional space-time. It was shown by Polyakov that compact planar QED exhibits charge confinement \cite{poly}. While the case of non-compact QED was studied by Grignani \emph{et. al.} \cite{qed3}.

In this paper, it is shown that an assumption of confinement of a certain
current gives rise to the photon mass.
The theory consider here consists of two species of free Dirac fermions living on
the plane, defined such that the theory is even under parity. These fermions are minimally coupled
to the photon field. It is found that
although the photons in the theory are massive, there is no spontaneous symmetry breaking.
It is also shown that when such a theory is defined over a manifold
with finite boundary, then there exist massless particles living on the
boundary.

In the following section the model is introduced and its various features are discussed. 
Section (\ref{sec2}) deals with the  effective action of photon and its mass.
Section (\ref{sec4}) deals with the case when the theory lives on a manifold with a finite boundary, followed by a brief summary.   

\section{\label{sec1}The model}

The Lagrangian describing two species of massive Dirac fermions living in 2+1 dimensional space-time reads:
\begin{equation} \label{lagone}
\mathscr{L}_{D} = \bar{\psi}_{+} ( i \gamma^{\mu}_{+}{\partial}_{\mu} - m  ) \psi_{+} + \bar{\psi}_{-} ( i \gamma^{\mu}_{-} {\partial}_{\mu} - m )\psi_{-}.
\end{equation}
Here, gamma matrices are defined for $\j_{+}$ field as $\g^{0}_{+} = \sigma_{3}, \g^{1}_{+} = i\sigma_{1} \text{  and  } \g^{2}_{+} = i\sigma_{2}$. Gamma matrices for $\j_{-}$ field are also same as $\j_{+}$ except for $\g^{2}$, which is defined as $\g^{2}_{+}=-\g^{2}_{-}$. This deliberate difference in choice of gamma matrices ensures that the Lagrangian is even under parity. It is known that, unlike four dimensional space-time, in the three dimensional world parity transformation is defined by reflecting one of the space axis, say Y axis, $(x,y) \rightarrow (x,-y)$. Instead of working with two spinor fields $\psi_{\pm}$, one can work in a reducible representation by defining $\Psi = (\psi_{+},\psi_{-})^{T}$, with $\beta = \gamma^{0} = \mathbf{1} \otimes \sigma_{3}$, $\alpha_{1}= \mathbf{1} \otimes \sigma_{1}$ and
$\alpha_{2}= \sigma_{3} \otimes \sigma_{2}$, so that the above Lagrangian now reads:
\begin{equation} \no
\mathscr{L}_{D} = \bar{\Psi} ( i \gamma^{\mu}{\partial}_{\mu} - m  ) \Psi,
\end{equation}    
where $\gamma_{1,2}=\beta \alpha_{1,2}$. Under parity operation, $\Psi$ transforms as
$\mathscr{P} \Psi(x,y,t) \mathscr{P}^{-1} = (\sigma_{1} \otimes \mathbf{1}) \Psi(x,-y,t)$. It can be checked that under this peculiar parity transformation, above Lagrangian remains even despite of having a mass term \cite{hagen}.

As it stands, apart from above mentioned parity transformation, the Lagrangian of this theory is invariant under two
independent continuous rigid transformations :
\begin{align} \label{sym1}
\j_{+}(r) \rightarrow e^{-i \q}\j_{+}(r), \,\j_{-}(r) \rightarrow e^{-i\q}
\j_{-}(r) ; \\ \label{sym2}
\j_{+}(r) \rightarrow e^{-i \l}\j_{+}(r), \,\j_{-}(r) \rightarrow e^{i \l}
\j_{-}(r) .
\end{align}
Here $\q$ and $\l$ are continuous real parameters. These being continuous symmetry operations, give rise to conserved currents as per the Noether theorem :
\be \no
\d_{\m} (j^{\m}_{+} + j^{\m}_{-}) = 0 \,\,\text{and}\,\, \d_{\m}(j^{\m}_{+} -
j^{\m}_{-})  = 0,
\ee
where $j^{\m}(r) = \bar{\psi}(r) \g^{\m} \psi(r)$. It turns out that under parity, current $J^{\m} = j^{\m}_{+} + j^{\m}_{-} = \bar{\J} \gamma^{\mu} \J$ transforms as a vector  \footnote{$\Lambda$ is diagonal matrix $\Lambda = \text{diag}(1,1,-1)$.}: $\mathscr{P} J^{\m}(x,y,t) \mathscr{P}^{-1} = \Lambda^{\m}_{\:\:\n} J^{\n}(x,-y,t)$, whereas $\tilde{J}^{\m} = j^{\m}_{+} - j^{\m}_{-} = - i \bar{\J} \gamma^{\mu} (\sigma_{3} \otimes \sigma_{3}) \J$, transforms as a pseudovector: $\mathscr{P} \tilde{J}^{\m}(x,y,t) \mathscr{P}^{-1} = \tilde{J}^{\m}(x,-y,t)$. Since the physical photon field transforms as a vector under parity operation: $\mathscr{P} A^{\m}(x,y,t) \mathscr{P}^{-1} = \Lambda^{\m}_{\:\:\n} A^{\n}(x,-y,t)$, its coupling with the current $j^{\m}_{+} + j^{\m}_{-}$ preserves parity while making the symmetry transformation (\ref{sym1}) local.

In this paper, we are interested in looking at the physical consequences if the current $j^{\m}_{+} - j^{\m}_{-}$ is confined. As pointed out by Kugo and Ojima in the context of QCD, and further discussed at length in Ref. \cite{nak}, that the statement of color charge confinement can be accurately stated as the absence of charge bearing states in the physical sector of the Hilbert space: $Q_{color} | phys \rangle = 0$. {In what follows, we shall work with a stronger condition than the Kugo-Ojima condition, and demand that the physical space of the theory described by Lagrangian (\ref{lagone}) should not have any states which carry $(j^{\m}_{+} - j^{\m}_{-})$ current, that is: $(j^{\m}_{+} - j^{\m}_{-}) | phys \rangle = 0$} \footnote{The physical space here stands for the set of states in the vector space of the theory, which do not have negative norm \cite{henn}. In case when the negative normed states are altogether absent, then the condition $(j^{\m}_{+} - j^{\m}_{-}) | phys \rangle = 0$ holds for the whole of Hilbert space and hence becomes an operator condition $(j^{\m}_{+} - j^{\m}_{-})= 0$.}. {This shall be referred to as \emph{current confinement condition} henceforth. Since we are demanding a priori that this current confinement condition should hold, it ought to be understood as a constraint. There exists a well known powerful technique to implement such a constraint using what is called the Lagrange multiplier (auxiliary) field \cite{henn}. One postulates the existence of a Lagrange multiplier field which is such that its only appearance in the action is via its coupling to the constraint condition. Thus the equation of motion corresponding to this field, obtained by demanding that the functional variation of the action with respect to this field be zero, is simply the constraint condition. It is worth pointing out that such Lagrange multiplier fields have no dynamics of their own, in the sense that there are no terms in the action comprising of spatial or temporal derivatives of these fields to begin with, and their sole purpose of existence is to ensure implementation of the constraint. Thus by enlarging the degree of freedom in the theory by an additional field, one ensures that the constraint condition gets neatly embedded into the action, and hence into the dynamics of the theory. In our case the Lagrange multiplier Bose field is $a_{\m}$, which is meant to implement the constraint $(j^{\m}_{+} - j^{\m}_{-})$, will only couple to it so that the Lagrangian (\ref{lagone}) gets an additional term}:  
\begin{equation} \label{lagtwo}
\mathscr{L} = \bar{\psi}_{+} ( i \gamma^{\mu}_{+}{\partial}_{\mu} - m  ) \psi_{+} + \bar{\psi}_{-} ( i \gamma^{\mu}_{-} {\partial}_{\mu} - m )\psi_{-} + a_{\m} \left( j^{\mu}_{+} - j^{\mu}_{-} \right). 
\end{equation}
{Note that the equation of motion for $a_\mu$ field: $\frac{\delta S}{\delta a_{\mu}} = 0$, gives the constraint $j^{\m}_{+} - j^{\m}_{-} = 0$.} Inorder to preserve parity, $a_\m$ field has to be a pseudovector owing to its coupling with pseudovector current. Thus $a_\m$ can in general be written as curl of some vector field $\chi$: $a_\m = \epsilon_{\m \n \l } \d^\n \chi^\l$, and can not have a contribution that can be written as a gradient of some scalar field. This asserts that $a_\m$ can not be a gauge field, since a gauge field under a gauge transformation transforms as a vector $\d_\m \Lambda$, which is not consistent with the pseudovector nature of $a_\m$. Further note that since $a_{\m}$ is curl of $\chi_\m$, it immediately follows that its divergence vanishes: $\partial_\m a^\m = 0$. 

%

In functional integral formulation of quantum field theory, generating functional is an object of central importance, which for this theory reads \footnote{Since the current in this theory couples directly to $a_\m$, it is treated as a dynamical variable instead of $\chi_\m$. Such a treatment is advocated by Hagen in Ref. \cite{hagen2}.}:
\begin{align}   \no
& {Z}[\e_{\pm}, \bar{\e}_{\pm}] = N \int \mathscr{D}[\bar{\psi}_{\pm}, {\psi}_{\pm}, a_{\m}]\, e^{i S}, \\ \no
&\text{where}\,\,\, S = \int d^{3}x\, \left( \mathscr{L} + \bar{\e}_{\pm} \j_{\pm} + \bar{\j}_{\pm} \e_{\pm} \right).
\end{align}
Here $\e$ and $\bar{\e}$ are external sources which are coupled to Fermi fields $\bar{\j}$ and $\j$ respectively. 

{Before we proceed with the details of the quantum theory, it is worth pointing out that if one functionally integrates $a_\mu$ in the above generating functional, one immediately obtains the current confinement condition $\delta(j^{\m}_{+} - j^{\m}_{-})$, since $a_\mu$ appears linearly in the action. This clearly shows that in the quantum theory as well, the Lagrange multiplier field $a_\mu$ is properly implementing the current confinement constraint.} 

Since the Lagrangian (\ref{lagtwo}) of the theory is invariant under transformation (\ref{sym2}), requirement that the generating functional of the theory should also be invariant under (\ref{sym2}), that is $\delta Z = 0$, is not unreasonable. Interestingly, it will be seen that this will give rise to Ward-Takahashi identities amongst various $n$-point functions in this theory and lead to non-trivial consequences. Demanding that $Z$ be invariant under infinitesimal version of transformation (\ref{sym2}) means $\del Z = 0$, which can be written as:
\begin{align} \no
\int \mathscr{D} [\bar{\psi}_{\pm},{\psi}_{\pm},{a}_{\mu}] \, (\delta S) \, e^{i S} = 0.
\end{align}
This can further be simplified to read:
\begin{align} \label{r1}
\int \mathscr{D} [\bar{\psi}_{\pm},{\psi}_{\pm},{a}_{\mu}] \, 
\left( \mp \bar{\e}_{\pm}(x) \psi_{\pm}(x) \pm \bar{\psi}_{\pm}(x) \e_{\pm}(x) \right)  \, e^{i S} = 0.
\end{align}
In terms of the generating functional of connected diagrams $W[\bar{\e}_{\pm},\e_{\pm}] = -i \ln Z[\bar{\e}_{\pm},\e_{\pm}]$,  equation (\ref{r1}) becomes:
\begin{align} \label{r2}
\bar{\e}_{+}(x) \frac{\delta W}{\delta \bar{\e}_{+}(x)} & - \bar{\e}_{-}(x) \frac{\delta W}{\delta \bar{\e}_{-}(x)}
\\ & \no - {\e}_{+}(x) \frac{\delta W}{\delta {\e}_{+}(x)} + {\e}_{-}(x) \frac{\delta W}{\delta {\e}_{-}(x)} = 0.
\end{align}
It is often convenient to work with the effective action $\Gamma[\bar{\psi}_{\pm},\psi_{\pm}]$ which is defined to be Legendre transform of $W[\bar{\e}_{\pm},\e_{\pm}]$: $W[\bar{\e}_{\pm},\e_{\pm}] = \Gamma[\bar{\psi}_{\pm},\psi_{\pm}] 
+ \int d^{3}x \, (\bar{\e}_{\pm} \psi_{\pm} + \bar{\psi}_{\pm} \e_{\pm} )$, so that equation (\ref{r2}) reads:
\begin{align} \label{master}
\frac{\delta \Gamma}{\delta \psi_{+}(x)} \psi_{+}(x) - & \frac{\delta \Gamma}{\delta \psi_{-}(x)} \psi_{-}(x) \\ \no
&- \frac{\delta \Gamma}{\delta \bar{\psi}_{+}(x)} \bar{\psi}_{+}(x) + \frac{\delta \Gamma}{\delta \bar{\psi}_{-}(x)} \bar{\psi}_{-}(x) = 0.
\end{align}
This is the master equation from which one can get Ward-Takahashi identities connecting various vertex functions, by taking appropriate derivatives. 
The two-point function for ${+}$ species of fermions $i S_{F}(x-y) = \langle T \left( \j_{+}(x) \bar{\j}_{+}(y)  \right) \rangle $, in terms of $\Gamma$ is given by: 
\begin{align} \no
S^{-1}_{F}(y-x) = \frac{\delta^{2} \Gamma}{\delta \psi(x) \delta \bar{\psi}(y)} \biggl \lvert_{\psi=\bar{\psi}=0.}
\end{align}
Taking functional derivative of the master equation (\ref{master}), once by $\bar{\psi}_{+}(y)$ followed by once with ${\psi}_{+}(z)$, one obtains following Ward-Takahashi identity for fermion Greens function:
\begin{align} \label{w1}
\del (x - z) S^{-1}_{F} (y-x) = \del (x - y) S^{-1}_{F} (x-z).
\end{align} 
This implies that $S^{-1}_{F} (y-x) = \del (x - y) \int d^{3}z \: S^{-1}_{F} (x-z)$, whose only solution is $S^{-1}_{F} (y-x) \propto \del (y-x)$. Above identity is very powerful, since it has allowed for an exact determination of propagator in this interacting theory. Exactly similar identity would also hold for propagator of ${-}$ species of fermions. It is worth mentioning, that this model is one of the rare cases where  full propagator of this theory is known without any approximation. Presence of a physically observable particle in a theory, manifests as poles of propagator in momentum space. In our case, as is clearly evident, the propagator is regular everywhere in momentum space, which means that the Dirac fermion in our theory is not a propagating mode. This is particularly surprising since we started with a free Dirac theory with a constraint condition on currents, and it appears that condition is severe enough to not allow free fermion propagation.

In the absence of Dirac fermions, it is a natural to inquire about the particle excitations in this theory. Inorder to answer this question, it is instructive to study the four-point function in this theory. Apart from a trivial non-propagating solution discussed above, assuming validity of translational invariance,  it can be shown that the Ward-Takahashi identity for four point function admits a solution of the kind:
$\langle T \left( \j_{+}(x_{1}) \j_{+}(x_{2}) \bar{\j}_{+}(y_{1}) \bar{\j}_{+}(y_{2})  \right) \rangle \propto \delta{(x_{1}-y_{1})}\,\delta{(x_{2}-y_{2})}\,f(x_{1}-x_{2})$, where $f$ is some function of $(x_{1}-x_{2})$. This means that this Ward-Takahashi identity allows for propagation of composite operator $\j(x) \bar{\j}(y) |_{x=y}$, which describes charge neutral excitations consisting of fermion-antifermion bound states.  It is worth mentioning, that the absence of fermions as elementary excitations and occurrence of bound states in a constrained theory like above, also appeared in a model of colour confinement proposed by Rajasekaran and Srinivasan \cite{sri}. Interestingly, they showed that quarks and gluons (which appeared as bound states) did not propagate and were confined, whereas mesons (colour neutral bound states of quarks) were propagating excitation in their model.

\section{\label{sec2}Electromagnetic response}

In this section we focus our attention on the electromagnetic response of the theory. Lagrangian (\ref{lagtwo}) with the minimal coupling of fermion fields to the photon field is given by:
\begin{align} 
\mathscr{L} & = \bar{\psi}_{+} ( i \slashed{\partial}_{+} - m + \slashed{a} + \slashed{A}  ) \psi_{+} + \bar{\psi}_{-} ( i \slashed{\partial}_{-} - m - \slashed{a} + \slashed{A} )\psi_{-} - \frac{1}{4} F_{\m \n} F^{\m \n}. \label{LagEM}
\end{align}
In order to find the response of the theory under the influence of photon field, it is imperative that the photon field be treated as an external field. Terms involving ghosts and gauge fixing, which are absent in the above Lagrangian, have been incorporated by appropriate modification of measure $\mathscr{D}[A_{\mu}]$. Inorder to take into account effects due to quantum corrections, which arise from virtual fermion loop excitation, one needs to find out the effective action by integrating out fermion field. The effective action upto quadratic terms in fields, obtained using derivative expansion of fermion determinant \cite{pkp1,das} reads:
\begin{equation} \label{lageff}
{\mathscr{L}}_{eff} = - \frac{1}{12 {\pi} |m|} {f_{\mu \nu}} {f^{\mu \nu}} - \frac{ m} {2 \pi |m|} \epsilon^{\mu \nu \rho} \left(  A_{\mu} {f}_{\nu \rho} + a_{\mu} {F}_{\nu \rho} \right) - \frac{1}{4} F_{\m \n} F^{\m \n}.
\end{equation}
It can be shown that, in the limit of large $m$ this approximation is valid and higher order terms can be neglected.
As is evident, $a_{\m}$ did not have a kinetic term to start with, but fermion loops have made it dynamical. Further, $A_\m$ and $a_\m$ fields are coupled by a mixed Chern-Simons term, which has a topological nature \cite{deser1,deser2,cs1,cs2,cs3}. In other words, this implies that $a_{\m}$ field has now become electromagnetically charged due to presence of virtual fermion cloud around it, with current being given by $J^{\mu}=\epsilon^{\mu \nu \rho} {\partial}_{\nu} {a}_{\rho}$, which is conserved off shell by construction. It is interesting to note that in this effective Lagrangian, the pseudovector field $a_\m$ is coupled to dual of $F_{\m \n}$ so that the effective action is even under parity.

{We started with a theory consisting of two species of massive Dirac fermions, with the assumption of current confinement. The current confinement being an independent condition, in the sense that it is not a consequence of the equations of motion of the theory, was understood as a constraint. We employed a judicious way of implementing this constraint using the Lagrange multiplier field $a_\mu$, which essentially does book keeping of the constraint. Even though constraint condition is stated in terms of fermion fields, it can not be viewed in isolation since the Dirac fields are coupled to the photon field. Thus even after integrating out the fermion fields, the effects of current confinement condition survive and manifest as coupling between $a_\mu$ and $A_\mu$ in (\ref{lageff}). Since the role of $a_\mu$ is only to ensure implementation of the constraint, it is imperative to integrate it out to see the effect of current confinement on the dynamics of the photon field. On integrating out $a_{\mu}$ field from  Lagrangian (\ref{lageff}), one arrives at an effective action for electromagnetic field}:
\begin{align} \label{eff}
\mathscr{L}_{eff} = - \frac{1}{4} F_{\m \n} F^{\m \n} + \frac{3 |m|}{\pi} F_{\m \n} \frac{1}{\d^{2}} F^{\m \n}.
\end{align}
As is evident, interaction with $a_{\m}$ field has induced gauge invariant mass $M = \tfrac{12 |m|}{\pi}$ for the physical electromagnetic field \cite{vsvyas}. One may wonder that the differential operator $\frac{1}{\partial^{2}}$ in the Lagrangian may compromise locality and causality. However it has been shown in Ref. \cite{vsvyas} that both of these features are intact.  The action of $\frac{1}{\partial^{2}}$ on a function becomes transparent by going over to Fourier space {\footnote{Alternatively, one may also consider the action of this differential operator in terms of convolution by a suitable Greens function $G(x)$, subject to the appropriate boundary conditions of the problem. The Greens function $G(x)$ is defined to solve: $\partial^{2} G(x) = \delta(x)$. With the knowledge of boundary conditions, formally this can be inverted: $G(x) = \frac{1}{\partial_{x}^{2}} \delta(x)$, so that $ \frac{1}{\partial_{x}^{2}} f(x) = \int dy  \frac{1}{\partial_{x}^{2}} \delta(x-y) f(y) = \int dy G(x-y) f(y)$.}, that is:
\begin{equation}
\frac{1}{\partial^{2}} f(x) = \int \frac{d^{3}p}{(2 \pi)^{3}} \frac{-1}{p^{2}} e^{- i p x} \tilde{f}(p),
\end{equation}
where $\tilde{f}(p)$ is Fourier transform of $f(x)$. Occurence of such terms in action have been long known, for example it is known to appear in the effective action of Schwinger model when one integrates out fermions, as also in the action of two dimensional gravity theory studied by Polyakov \cite{vsvyas}.

It is known that in the planar world there exists Chern-Simons Lagrangian: 
\begin{equation}
\mathscr{L} = - \frac{1}{4} F_{\m \n} F^{\m \n} + \frac{M}{2} \epsilon_{\m \n \l} A^{\m} \partial^{\n} A^{\l}
\end{equation}
which is also gauge invariant and describes massive photon field \cite{deser2}. However unlike Lagrangian (\ref{eff}) the mass term in this theory has a topological origin, and the theory evidently violates parity.

It may be tempting to believe that the photon mass term in the theory occurs because of some kind of spontaneous symmetry breaking and associated Anderson-Higgs mechanism. However, note that the theory is invariant under two kinds of continuous rigid transformations, which are generated by two conserved charges $Q_{1,2} = \int d^{2}x : j^{0}_{+} \pm j^{0}_{-}:$. Vacuum expectation value of conserved current $\lan vac | j^{\mu}_{\pm}(x)| vac \ran$ can be written as \cite{fujikawa}:
\begin{align} \no
\lan j^{\mu}_{\pm} \ran & = \lim_{x \to y} \lan \bar{\psi}_{a,\pm}(y) \gamma^{\mu}_{ab,\pm} \psi_{b,\pm}(x) \ran  \\
& = - i \lim_{x \to y} \text{tr} \: \gamma^{\mu}_{\pm} \, S_{F,\pm}(x-y),
\end{align}
where \emph{tr} stands for trace over Dirac indicies. Since $S_{F,\pm}(x-y) = \text{const.} \times \delta (x-y)$ in this theory, one finds that $\lan vac | j^{\mu}_{\pm} | vac \ran = 0$. This straightforwardly implies that, 
the charges $Q_{1,2}$ annihilate the vacuum  $Q_{1,2} |vac \ran = 0$ in this theory. This emphatically shows that there is no spontaneous symmetry breaking whatsoever, and that the current confinement is responsible for the photon mass.

\section{\label{sec4}Boundary theory}

In above discussions we have assumed that the theory lives on two dimensional manifold whose boundary lies at the infinity, and further all the fields in discussion were assumed to decay sufficiently quickly so that surface terms in the action contribute negligibly. In this section we shall consider the case when the boundary is finite, in which case it may not be possible to ignore contribution due to the surface terms.

As noted above, low energy effective action describing the dynamics of low energy electronic excitation, subject to the current constraint, coupled to electromagnetic field is given by (\ref{lageff}):
\begin{align} \nonumber
\mathscr{L} = - \frac{1}{4} F_{\m \n} F^{\m \n} - \frac{1}{12 {\pi} |m|} {f_{\mu \nu}} {f^{\mu \nu}} - \frac{ m} {2 \pi |m|} \epsilon^{\mu \nu \rho} \left(  A_{\mu} {f}_{\nu \rho} + a_{\mu} {F}_{\nu \rho} \right).
\end{align}
Note that the last mixed Chern-Simons term is not invariant under local gauge transformation:
$A_{\mu} \rightarrow A_{\mu} + \d_{\mu} \Lambda$, where $\Lambda$ is some regular function of $x$. As a result, the change in action is given by:
\begin{equation} \no
\delta S_{CS} = \left( \frac{sgn(m)}{2 \pi} \right) \int d^{3}x \,\, \epsilon^{\m \n \r} \d_{\m}
 \left( \Lambda \, f_{\n \r} \right).
\end{equation}
Above volume integral can be converted to a surface integral, defined on closed boundary of the manifold, to give an action:
\begin{equation} \no
\delta S_{CS} = \left( \frac{sgn(m)}{2 \pi} \right) \int_{B} d^{2}x \,\, \epsilon^{\m \n } \Lambda \, f_{\m \n}.
\end{equation}
This term, as it stands, is not gauge invariant, and is defined on the boundary, which encloses the bulk. Gauge invariance of any given theory, is a statement that, the theory is constrained, and possesses redundant variables. We observe that, our theory to start with was gauge invariant at classical level. One loop corrections arising out of fermion loops, generate Chern-Simons term, which exhibits gauge noninvariance. Because, our theory to start with was gauge invariant, and hence constrained, consistency demands that quantum(corrected) theory should also respect the imposed constraints, and hence should be gauge invariant. The occurrence of above gauge noninvariance, simply implies that one is only looking at one particular sector of the theory, and there exists other dynamical sector, whose dynamics is such that it compensates with the one above to render the total theory gauge invariance. Following Ref. \cite{wilczekbook}, we demand that there must exist a corresponding gauge theory living on the boundary, defined such that it contributes a gauge noninvariant term of exactly opposite character and hence cancels the one written above. The simplest term, living on boundary, that obeys above condition is:
\begin{equation} \no
S_{B} = \frac{-sgn(m)}{2 \pi} \int_{B} d^{2}x \, \theta \epsilon^{\m \n} f_{\m \n},
\end{equation}
where $\theta(x,t)$ is a Bose field, which transforms as $\theta \rightarrow \theta + \Lambda$ under a gauge transformation. In general, this scalar field would be dynamical, and with a gauge invariant kinetic term, the boundary action reads:
\begin{equation} \no
S_{B} = \int_{B}d^{2}x \,\, \left[ c \,\left( \d_{\m} \theta - A_{\m} \right)^{2} - \frac{sgn(m)}{2 \pi} \theta \epsilon^{\m \n} f_{\m \n} \right].
\end{equation}
Owing to its peculiar transformation gauge transformation property, a quadratic mass term for $\theta$ is not gauge invariant. Hence, in a gauge theory framework like this, $\theta$ field remains massless. 
Since the coupling of $\theta$ field with $a$ field is anomalous, it turns out that the chiral current in this quantum theory is no longer conserved.

\section{\label{sec5}Conclusion}

In this paper, we have shown that, in a parity invariant theory of two free Dirac fields living on a plane, confinement of current $j^{\m}_{+}-j^{\m}_{-}$ gives rise to the photon mass. To the best of our knowledge this is the only model in which current confinement paves the way to the gauge boson mass. A unique feature of this mechanism of photon mass generation is that there is no kind of spontaneous symmetry breaking involved. It is found that in case when the theory is defined on a manifold with a boundary, consistency implies the existence of massless particles on the boundary. It would be interesting to investigate whether it is possible to have a composite photon from confinement, as was seen in planar QED with a tree level Chern-Simons term \cite{abhi}. Further, it is believed that the connection between gauge boson mass, compositeness and current confinement, as seen in this theory, could have some implications in the theory of strong interactions - QCD.   
Work along these lines is in progress and shall be published in due course.

\section*{Acknowledgements}
The authors would like to thank the anonymous referee for his/her constructive comments which has helped the authors again a better understanding of this theory. The authors also thank Prof. V. Srinivasan for several useful discussions.  




%
\providecommand{\noopsort}[1]{}\providecommand{\singleletter}[1]{\#1}%

\end{document}